# An Alternative to Multi-Factor Authentication with a Triple-Identity Authentication Scheme

Suyun Borjigin
Independent Researcher, China
yunsu000@hotmail.com

**Abstract --** The existing authentication system has two entry points (i.e., username and password fields) to interact with the outside, but neither of them has a gatekeeper, making the system vulnerable to cyberattacks. In order to ensure the authentication security, the system sets a third entry point and use an external MFA service to guard it.

The crux of the problem is that the system has no internal mechanism to guard its own entry points as no identifiers can be defined for the username and password without using any personal information. To solve this problem, we open the hash algorithm of a dual-password login-authentication system to three login credentials. Therefore, the intermediate elements of the algorithm can be used to define an identifier to verify the user identity at each entry point of the system.

As a result of the above setup, a triple-identity authentication is established, the key of which is that the readily available user's login name and password are randomly converted into a matrix of meaningless hash elements which are concealed, incommunicable, inaccessible, and independent of personal information. So the identifiers defined using such elements can be used by the system to verify the identities of the user at all the entry points of the system, thereby ensuring the authentication security without relying on MFA services.

## 1 INTRODUCTION

Today, almost all online services have enabled Multi-Factor Authentication (MFA) [1, 2] which is a third-party service and used to help verify the identity of a user using an extra code sent to the user by the MFA service. Typically, the code is either transmitted through the network or generated by an authenticator app on their smartphone and then manually entered by the user.

A user authentication system usually involves three login credentials, i.e. a username, a password, and a hash value which is generated from the password by a hash algorithm. In the prior arts, however, only one of them (usually the hash value) is associated with the identity of the user and then verified on the server, while the username and password are not uniquely represented by an identity. This means that the system cannot verify the identities of the user at the entry points (i.e., the username and password fields), resulting in no gatekeepers protecting these entries. In addition, in a system enabling MFA, the extra code transmitted through the network or created from another app has the potential to be intercepted by hackers, while the manual input of it is vulnerable to certain types of malwares [3, 4, 5]. Particularly, due to the lack of identity resources of a user, the prior arts often use the IMEI or IMSI number of the user's smartphone as the identifier to prove their identity. However, the user of personal information as an identifier has proven to be a bad idea because it is readily available to the public.

If truly secret identifiers can be defined for the login name and the password, all the entry points of the system can be protected by gatekeepers. If the identifiers can be managed completely by the system and no longer involve any personal information owned by the user, not only is the authentication secure but each identifier is also unbreakable. And if the online transmission and the manual input of the MFA code can be replaced by an internal mechanism of the system, such an exclusive system in itself is sufficient to protect the authentication without relying on external services.

To achieve these goals, a triple-identity authentication is established based on a dual-password login-authentication system, in which a login password entered by a user via their subscribed smartphone is converted into an authentication password by a hash algorithm referred to as a quasi-matrix password converter [6-8], as shown in Figure 1. The key advantage of this authentication scheme is that the randomly generated intermediate elements of the hash algorithm (or the converter) are ideal for the system to generate truly secret identifiers to verify the identities of the login name and the login password. Accordingly, we open the hash algorithm to all login credentials, making it an open hash algorithm.

More importantly, in this authentication scheme, personal information (such as the user's name, address, username, phone number, password, IMEI, IMSI, or any information owned by the user) will only be used to define the identities of the user. Instead, the corresponding identifiers of the user must be generated with the meaningless intermediate hash elements completely managed by the system.

During registration, the login name (i.e., the username of the user's email address and the phone number) is converted by the hash algorithm to generate a LN converter, as shown in Figure 2. Then, the system can randomly select some hash elements from the converter to define a LN identifier. As the use of a single factor to define the user identity is risky, the triple-identity authentication system combines several factors (i.e., the login name, the IMEI and IMSI numbers of the subscribed smartphone of the user) to define a LN identity of the user. Then, the system associates the user's combined LN identity with the system-generated LN identifier.

Next, the user-entered login password (LP) is converted to generate a LP converter and an authentication password (AP), as shown in Figure 1. Then, the system randomly selects some hash elements from the converter to define a LP identifier. And the system combines the login password, the IMEI and IMSI numbers of the subscribed smartphone into a LP identity of the user, which can then be associated with the system-generated LP identifier.

Finally, for the user verification on the server, the system defines the authentication password as an AP identifier, combines the IMEI and IMSI numbers of the subscribed smartphone of the user to define the AP identity of the user, and then associates it with the AP identifier.

Consequently, a triple-identity authentication is established based on the dual-password login-authentication system, by which the combined identities of the user can be verified at the two entry points and on the system server. Particularly, the system-managed LN and LP identifiers do not need to be transmitted through the network and cannot be accepted by the login fields (i.e., the LN and LP fields) as they contain invalid characters that are forbidden by the fields.

## 2 QUASI-MATRIX PASSWORD CONVERTER

The dual-password login-authentication system is created by integrating a pair of login and authentication passwords [7, 8], in which a login password entered by a user through their subscribed smartphone is converted into an authentication password by the hash algorithm or quasi-matrix password converter, as shown in Figure 1.

After every login character is converted into a string by the hash algorithm based on the selected character digit, all the strings are shuffled together to generate a quasi-matrix password converter and then an authentication password according to the special rules of the shuffling labels.

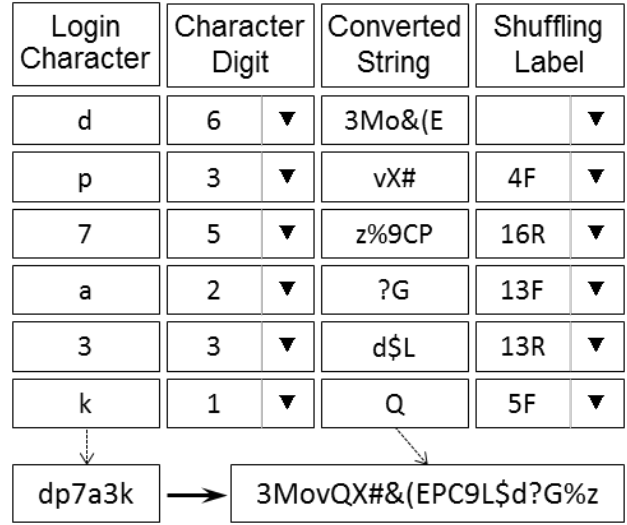

Figure 1: The quasi-matrix password converter

Clearly, the hash algorithm runs in the system background without the involvement of users. That is, the intermediate hash elements (excluding the login character column) are concealed, incommunicable, inaccessible, and independent of any personal information readily available to the public. This suggests that the characters in any languages that a computer can process can be used in the composition of the converted strings, that is, the authentication password and the identifiers.

The elements with such properties are ideal for the system to define truly secret and unbreakable identifiers for the three login credentials (that is, a login name, a login password, and an authentication password). To take full advantage of such properties, the hash algorithm or the quasi-matrix password converter of the dual-password login-authentication system should be open to all the login credentials. Therefore, the meaningless hash elements can be selected by the system to define the identifiers and verify the identities of the user at each entry point of the system. This means that the security of the authentication system can eventually be guarded by an internal gatekeeper mechanism rather than an external service.

There are multiple ways for the system to select the hash elements to generate an identifier. An element row including the login character, an element column excluding the Login Character column, or a combination of some hash elements randomly selected from the converter can be truly secret. So, any one of the rows, columns, or combinations can be set up as an identifier and associated with the corresponding login credential by the system.

Furthermore, based on the converter structure, the pair of login and authentication passwords can be managed by the user and the system separately without affecting each other [6-8]. Thus, they can be set completely different in password strength for both parties to satisfy their needs for usability and security, and also for the password field to distinguish between the passwords. The login password can be set by the user within a specific range in length from five to fifteen characters and only contain lowercase letters and/or digits. On the contrary, the authentication password may be set to be twenty characters in length and must contain at least four character classes (namely uppercase and lowercase letters, digits, and symbols) [9, 11], and the first four characters of each authentication password must contain either one uppercase letter or one symbol. In the triple-identity system, characters other than lowercase letters and digits are considered invalid and forbidden by the login fields. So the LN and LP fields will reject authentication passwords and identifiers for containing invalid characters.

## 3 REGISTRATION OF THE LOGIN CREDENTIALS

A traditional user authentication system involves three login credentials, but only one of them (usually the hash value) is uniquely represented by an identity, while the others are not. And the only identity is verified using an identifier containing personal information, making it difficult to ensure the security of the authentication.

In the triple-identity authentication, with the hash algorithm open to the login credentials, the readily available login name or the login password can be converted into a matrix of truly random elements. Therefore, the identifiers for the identity verification at the entry points of the system can only be generated using the random hash elements. And the user's identities can only be combined with personal information owned by the user. In the following sections, the associations between the identities and identifiers of the user will be established and registered on the server.

### 3.1 Registration of the login name

When registering the login name, the user provides their personal information as required through their subscribed smartphone. The system also collects the identification information of their smartphone, such as the phone number, International Mobile Equipment Identity (IMEI) number, and International Mobile Subscription Identity (IMSI) number.

In view of the openness of the hash algorithm to the login credentials, either the username of the email address or the phone number can be converted by the hash algorithm to generate a login name (LN) converter. In Figure 2, the username is converted. So the login name is the username.

There is no need to shuffle the converted strings together to generate a string like an authentication password. As with the login password, the login name is also five to fifteen characters long and contains lowercase letters and/or digits. Uppercase letters will be turned into lowercase, if any, while other characters will be removed by the system.

Subsequently, the system randomly selects some hash elements from the LN converter to define a LN identifier and associate it with the user's LN identity, which is defined as the combination of the username of the email address or the phone number, the IMEI and IMSI numbers of the subscribed smartphone. Thus, the system has a gatekeeper to protect the LN entry point.

| Username | Character Digit | Converted String | Shuffling Label |
|---|---|---|---|
| B | 3 ▼ | y]Q | ▼ |
| e | 5 ▼ | #ws%8 | 5F ▼ |
| n | 3 ▼ | O^& | 9R ▼ |
| z | 2 ▼ | $d | 17R ▼ |
| 4 | 3 ▼ | )Lh | 13F ▼ |
| 2 | 3 ▼ | zF= | 8F ▼ |
| 8 | 1 ▼ | m | 11F ▼ |

Figure 2: The login name converter of the username

Specifically, the system may associate the user's LN identity "Benz428+IMEI+IMSI" with the LN identifier "4O^&17R2zF=", as shown in Figure 2. The former is the combination of the username of the email address Benz428@woxinet.com, the IMEI and IMSI numbers of the smartphone, which can be used to symbolize the process of entering the username via the smartphone. And the latter is the LN identifier generated by combining the elements "4", "O^&", "17R", "2", and "zF=" randomly selected from the LN converter by the system.

### 3.2 Registration of the login password

When registering the login password, a quasi-matrix password converter (i.e., the LP converter) is generated once the length and password characters are determined by the user, as shown in Figure 1. Then, the system can randomly select some hash elements from the converter to define a LP identifier and associate it with the LP identity of the user, which is defined as the combination of the login password, the IMES and IMSI numbers of the subscribed smartphone. Thus, the system has a gatekeeper to protect the LP entry point.

Specifically, the system may associate the user's LP identity "dp7a3k+IMEI+IMSI" with the LP identifier "z%9CP213Rp", as shown in Figure 1. The former is the combination of the login password "dp7a3k", the IMEI and IMSI numbers of the smartphone, which can be used to symbolize the process of entering the login password through the smartphone. And the latter is the LP identifier generated by combining the elements "z%9CP", "2", "13R", and "p" randomly selected from the LP converter by the system.

### 3.3 Registration of the authentication password

When registering the authentication password (AP), it can be generated by the LP converter in Figure 1. The system can then define an AP identifier with the authentication password alone and associate it with the AP identity of the user, which is defined as the combination of the IMEI and IMSI numbers of the smartphone. Optionally, the combination can also be generated in the manner as described in the sections 3.1 and 3.2, and associated with the AP identifier. As a result, the system can verify the AP identity on the server and grant the user access to their account.

## 4 VERIFICATION OF THE LOGIN IDENTITIES

An authentication system interacts with the outside mainly through two entry points, i.e., the username and password fields at the user interfaces, and hackers also utilize these entries to launch cyberattacks on the authentication system. With the triple-identity authentication in place, the combined identities of the user will be verified at these entries with the identifiers generated using the truly random hash elements.

When accessing the login name page via a smartphone, in response to the access or the input on this page, the system checks the database for a matching LN identifier. For the triple-identity system, it is easy to identify if the combined LN identity "LN+IMEI+IMSI" matches a LN identifier stored in the database as long as the user has registered. Then, the system grants the user access to the password page. Otherwise, the access is rejected.

If the user can proceed to the password page, in response to the password input through their smartphone, the system checks the database for a matching LP identifier. If the combined LP identity "LP+IMEI+IMSI" matches a LP identifier, the system can grant access to the final stage of the authentication. Otherwise, the access will be rejected.

At the final stage of the authentication, the system verifies the combined AP identity "IMEI+IMSI" with the AP identifier (i.e., the system-generated authentication password). After the previous successful LN and LP identity verifications, the system can definitely find the match and then grant the user access to their account.

## 5 CONCLUSION

This is a new authentication system that sets a gatekeeper mechanism to protect the security of the authentication at all entry points that the system interacts with the outside.

As highlighted in this paper, existing authentication systems need to address their own problems so as to no longer rely on external protection. Here we proposed a triple-identity authentication for dealing with the systematic problems. In this authentication, the text password is still essential, but it is no longer a core factor. Instead, a gatekeeper mechanism that can verify the user's identities at all entry points of the system becomes the key to solving the problems.

By opening the hash algorithm, the readily available login name and login password can be randomly converted into meaningless elements in order to define the identifiers, while any personal information only needs to be used to define the identities of the user. And the configuration of password strength and login fields makes the credentials, identities, and identifiers useless on any unregistered devices as they contain invalid characters. Accordingly, the gatekeeper can implement pairwise verification of the user's identity versus their identifier at every entry point of the system, meaning that it can sufficiently and effectively control the interactions between the user and the system.